\documentstyle[prl,aps,epsfig,floats]{revtex}

\newcommand\beq{\begin{equation}}
\newcommand\eeq{\end{equation}}
\newcommand\bea{\begin{eqnarray}}
\newcommand\eea{\end{eqnarray}}
\newcommand\nonu{\nonumber}
\newcommand\ua{\uparrow}
\newcommand\da{\downarrow}

\begin{document}

\draft

\textheight=23.8cm
\twocolumn[\hsize\textwidth\columnwidth\hsize\csname@twocolumnfalse\endcsname

\title{\Large Junction of several weakly interacting quantum wires: a 
renormalization group study}
\author{\bf Siddhartha Lal$^1$, Sumathi Rao$^2$ and Diptiman Sen$^1$} 
\address{\it $^1$ Centre for Theoretical Studies,
Indian Institute of Science, Bangalore 560012, India \\ 
$^2$ Department of Physics, Pennsylvania State University, State College, PA
16802, USA \\
and Harish-Chandra Research Institute, Chhatnag Road, Jhusi, Allahabad 
211019, India}

\date{\today}
\maketitle

\begin{abstract}
We study the conductance of three or more semi-infinite wires which meet at a 
junction. The electrons in the wires are taken to interact weakly with each 
other through a short-range density-density interaction, and they encounter a 
general scattering matrix at the junction. We derive the renormalization group
equations satisfied by the $S$-matrix, and we identify its fixed points and
their stabilities. The conductance between 
any pair of wires is then studied as a function of physical
parameters such as temperature. We discuss the possibility of observing the 
effects of junctions in present day experiments, such as the four-terminal 
conductance of a quantum wire and crossed quantum wires. 
\end{abstract}
\vskip .5 true cm

\vskip.5pc
]
\vskip .5 true cm

\section{\bf Introduction}

Recent advances in the fabrication of semiconductor heterostructures have
made it possible to study electronic transport in a variety of geometries.
Recent studies of ballistic transport through a quantum wire (QW) have brought
out the important role played by both scattering centers and the interactions 
between the electrons inside the QW. Theoretical studies using a 
renormalization group (RG) analysis show that repulsive interactions between 
electrons tend to 
increase the effective strength of the scattering as one goes to longer
distance scales \cite{kane}; experimentally, this leads to a decrease in the 
conductance as the temperature is reduced or the wire length is increased 
\cite{tarucha}. Considerable effort has also gone into understanding the 
effects of (Fermi) leads \cite{safi}, multiple impurities \cite{maslov} 
and also contacts \cite{lal}. Motivated by the understanding of the 
effects of scattering
in a one-dimensional problem, we are led to address the following question 
in this work: what is the effect of interactions between electrons on the 
conductance of more complicated geometrical structures such as three or 
more QWs meeting at a junction? This problem has been studied before in Ref. 
\cite{nayak}; as explained below, our model differs from theirs in certain 
ways, and our results are quite different. We will show that for the case of 
weak interactions, the effects of a junction (characterized by an arbitrary 
scattering matrix $S$) on the conductance can be understood in great detail 
by using a RG technique introduced in Ref. \cite{yue}. We will also 
complement this with a study of the effects of certain special kinds of 
junctions for arbitrary electron interaction to gain a more complete picture.

The plan of the paper is as follows. In Sec. II, we will define a junction 
in terms of a scattering matrix, and we will provide a microscopic lattice
model of a junction. In Sec. III, we will discuss an interacting theory of 
spinless fermions in the presence of an $S$-matrix at the junction, and we 
will enumerate some of the special $S$-matrices for which the theory can be 
bosonized. Sec. IV will contain a derivation of the RG equations for 
the junction 
$S$-matrix for the case of weak interactions in the wires. In Sec. V, we will 
study the fixed points of the RG equations and their stabilities for the case 
of three wires meeting at a junction. Wherever possible, we will compare our 
weak interaction results with the exact results available from bosonization.
In Sec. VI, the results of the previous section will be used to study the
conductance of a three-wire system as a function of the temperature in 
the vicinity of one of the fixed points. In Sec. VII, we will consider
the temperature dependence of the four-terminal conductance of a quantum 
wire (which is often studied experimentally). In Sec. VIII, we will study the
fixed points and stabilities of the RG equations of a four-wire system, and 
its four-terminal conductance. In Sec. IX, we will briefly discuss how to
extend the previous analysis to the case of spinful fermions. We will make
some concluding remarks (including a comparison of our model to that given
in Ref. \cite{nayak}) in Sec. X.

\section{\bf A Model For The Junction}

To study the problem, we first need a model for the junction. Let us assume 
that $N$ semi-infinite wires meet at a junction. The wires are parameterized 
by coordinates $x_i$, $i=1,2,...,N$. The junction is the point where all the 
$x_i$ are simultaneously equal to 0. We adopt the convention that each $x_i$ 
increases from $0$ as one goes outwards from the junction along wire $i$.
Let us denote the incoming and outgoing one-electron wave functions on wire 
$i$ by $\psi_{Ii} (x_i)$ and $\psi_{Oi} (x_i)$ respectively (we are ignoring
the spin label $\sigma$ for the moment); see Fig. 1. For a given wave number 
$k>0$, these wave functions are proportional to the plane waves $\exp 
(-ikx_i)$ and $\exp (ikx_i)$. 

\vspace*{-.4 cm}
\begin{figure}[htb]
\begin{center}
\epsfig{figure=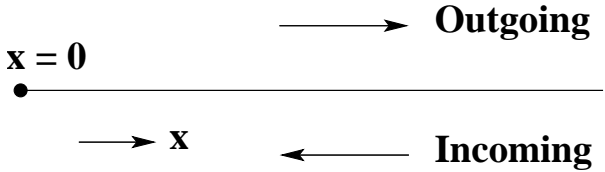,width=8.5cm}
\end{center}
\caption{Picture of a single wire showing the incoming and outgoing directions
and the junction at $x=0$.}
\end{figure}

The coefficients of the plane waves are
related to each other by a $N \times N$ scattering matrix called $S$. 
Denoting the incoming and outgoing wave functions at the junction by the 
columns $\psi_I (0)$ and $\psi_O (0)$, we have the relation
\beq
\psi_O (0) ~=~ S ~\psi_I (0) ~.
\eeq
Clearly, $S$ must be unitary. (If we want the system to be invariant under 
time reversal, $S$ must also be symmetric). The diagonal entries of $S$ are 
the reflection amplitudes $r_{ii}$, while the off-diagonal entries are the 
transmission amplitudes $t_{ij}$ to go from wire $j$ to wire $i$.

It is useful, though not necessary, to have in mind a microscopic model
of a junction with an unitary and symmetric
$S$-matrix. A simple lattice model for this is shown
in Fig. 2 for the case of three wires labeled by $i=1,2,3$. The junction is 
the site labeled as 0, while the sites on the wires have labels going from 
1 to $\infty$. The electrons hop from site to site with a hopping constant
which is $-1$ on all bonds except the 3 bonds which join the sites labeled as 
1 with the junction; on those three bonds, we take the hopping constants to
be the real numbers $-u_i$. In addition, we have a chemical potential $\lambda$
at the junction, while the chemical potential on all the other sites is $0$.
The momenta of the electrons go from $-\pi$ to $\pi$ (taking the lattice
spacing to be 1); the dispersion relation is given by $E = - 2 \cos k$.
Since the chemical potential is zero at all sites except one, the system is at
half-filling, and the Fermi points lie at $\pm k_F$ where $k_F =\pi /2$.
For incoming momenta $k$ close to $k_F$, we find that the entries
of the $S$-matrix are given by
\bea
r_{ii} ~&=&~ \frac{2u_i^2}{D} - 1 ~, \nonu \\ 
t_{ij} ~&=&~ \frac{2u_i u_j}{D} ~, \nonu \\
{\rm where} \quad D ~&=&~ \sum_{k=1}^3 ~u_k^2 ~+~ i \lambda ~.
\label{latsmat}
\eea
This matrix is both unitary and symmetric, although it is not the most
general possible matrix with those properties.

\vspace*{-.4 cm}
\begin{figure}[htb]
\begin{center}
\epsfig{figure=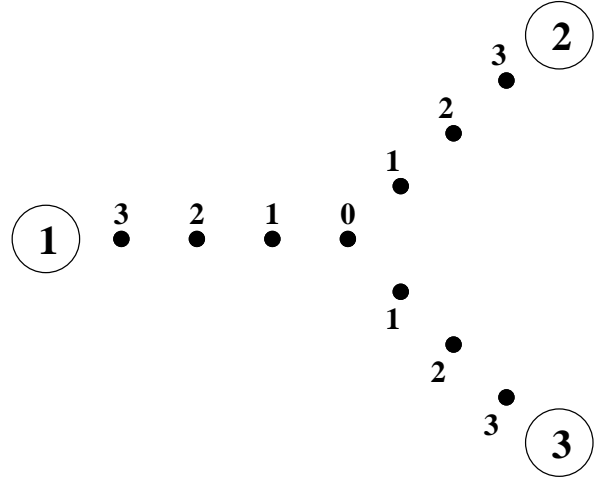,width=8.5cm}
\end{center}
\caption{Picture of the lattice model for three wires meeting at a junction.}
\end{figure}

\section{\bf Tomonaga-Luttinger Liquids Without Spin}

Let us now discuss the fermion fields in some more detail. We will consider a 
single wire for the moment, so that the label $i$ can be dropped. Since all 
low-energy and long-wavelength processes are dominated by modes near the 
Fermi points $\pm k_F$, let us write the second-quantized field $\Psi (x)$ 
(which implicitly contains both fermion annihilation operators and wave 
functions) as 
\beq
\Psi (x) ~=~ \Psi_I (x) ~e^{-ik_F x} ~+~ \Psi_O (x) ~e^{ik_F x} ~. 
\label{psiio}
\eeq
Note that the fields $\Psi_I$ and $\Psi_O$ defined in Eq. (\ref{psiio}) vary 
slowly on the scale of the inverse Fermi momentum $k_F^{-1}$, since we have 
separated out the rapidly varying functions $\exp (\pm i k_F x)$. We will
henceforth use the notation $\Psi_I$ and $\Psi_O$ for these slowly varying
second-quantized fields, rather than the incoming and outgoing fields defined 
earlier. For these fields, we will only be interested in Fourier components 
with momenta $k$ which satisfy $|k| << k_F$. We can then make a linear 
approximation for the dispersion relations which take the form $E= \pm 
\hbar v_F k$ for the fields $\Psi_O$ and $\Psi_I$ respectively, where $v_F$
is the Fermi velocity. (For instance, for the lattice model 
discussed above, $v_F = 2 \sin k_F$). We will also assume that the entries
of the $S$-matrix do not vary with $k$ in the limited range of momenta that
we are interested in.

We now introduce a model for interactions between electrons. Let us consider 
a short-range density-density interaction of the form
\beq
H_{\rm int} ~=~ \frac{1}{2} ~\int \int dx dy ~\rho (x) ~V (x-y) ~\rho (y) ~,
\label{hint1}
\eeq
where $V(x)$ is a real and even function of $x$, and the density $\rho$ is 
given in terms of the fermion field as $\rho (x) = \Psi^\dagger (x) \Psi (x)$.
Using Eq. (\ref{psiio}), we find that
\bea
\rho (x) ~&=&~ \Psi_I^\dagger \Psi_I ~+~ \Psi_O^\dagger \Psi_O \nonu \\
& & ~+~ \Psi_I^\dagger \Psi_O ~e^{i2k_F x} ~+~ \Psi_O^\dagger \Psi_I ~
e^{-i2k_F x} ~.
\label{dens}
\eea
We can now rewrite the interaction in Eq. (\ref{hint1}) in a simple way 
if $V(x)$ is so short-ranged that the arguments $x$ and $y$ of the two 
density fields can be set equal to each other wherever possible. [In doing
so, we will be ignoring terms which have scaling dimension greater than 2, and
are therefore irrelevant in the RG sense. We note that the assumption of a 
short-ranged interaction is often made in the context of the Tomonaga-Luttinger
liquid description of systems of interacting fermions in one dimension.] Using
the anticommutation relations between different fermion fields, we obtain
\beq
H_{\rm int} ~=~ g_2 ~\int dx ~\Psi_I^\dagger \Psi_I \Psi_O^\dagger \Psi_O ~,
\label{hint2}
\eeq
where $g_2$ is related to the Fourier transform of $V(x)$ as $g_2 = {\tilde V}
(0)- {\tilde V} (2k_F)$. [Note that $g_2$ is zero if $V(x)$ is a 
$\delta$-function; so $V(x)$ should have a finite range in order to have an
effect.] Thus the interaction depends on a single parameter $g_2$ on each wire.
Different wires may have different values of this parameter which we will 
denote by $g_{2i}$. For later use, we define the dimensionless constants
\beq
\alpha_i ~=~ \frac{g_{2i}}{2\pi \hbar v_F} ~,
\label{ali}
\eeq
where we assume that the velocity $v_F$ is the same on all wires.

For many problems involving a Tomonaga-Luttinger liquid, it is useful to 
bosonize the theory \cite{rao,gogolin}. For spinless fermions, the bosonic 
theory is characterized 
by two quantities, namely, the velocity of the bosonic excitations
$v$, and a dimensionless parameter $K$ which is a measure of the interactions 
between the fermions. (Typically, $K$ governs the exponents which appear in 
the power-law fall-offs of various correlation functions in the theory). For 
a model defined on the entire real line with the interaction parameter $g_2$ 
or $\alpha$ defined above, we find that \cite{rao}
\bea
v ~&=&~ v_F ~(1 - \alpha^2 )^{1/2} ~, \nonu \\
K ~&=&~ \bigl( \frac{1-\alpha}{1+\alpha} \bigr)^{1/2} ~.
\eea 
Thus $K=1$ for noninteracting fermions, while $K<1$ for short-range repulsive 
interactions. For weak interactions, we see that $v=v_F$ while $K = 1 -\alpha$
to first order in $\alpha$. In this work, we will be interested in the
case in which the interactions are weak and repulsive, i.e., the parameters 
$\alpha_i$ are all positive and small.

Although bosonization is a very powerful technique, it is not always possible 
to bosonize a system of interacting fermions with boundaries. In particular, 
for our system of interest, i.e., three or more semi-infinite wires meeting 
at a junction with some arbitrary $S$-matrix defined at that point, 
bosonization is a difficult task in general. The reason is that although one
can always find linear combinations of the incoming and outgoing fermion fields
which unitarily diagonalize the $S$-matrix, the four-fermion interactions
in the bulk of the wires are generally not diagonal in terms of the same linear
combinations. Conversely, the interactions in the bulk of the wires can be
bosonized, but it is then generally not clear what boundary conditions 
should be imposed on the bosonic fields at the junction.

However, it is possible to bosonize the system easily for some special 
forms of the $S$-matrix at the junction. For the case of three 
wires, there seem to be only six such forms. These are as follows. 
 
\noindent (a) Case I: Here $|r_{11}|=|r_{22}|=|r_{33}|=1$, and all the
other entries of the $S$-matrix are zero. This can be realized by the lattice 
model of Fig. 2 if we take the limit $\lambda \rightarrow \infty$. This case
corresponds to the three wires being disconnected from each other. Each 
wire can then be bosonized by an unfolding technique described in 
Ref. \cite{gogolin}. 

\noindent (b) Cases II-IV: In case II, $|r_{33}|=|t_{12}|=|t_{21}|=1$, and all
the other entries of $S$ are zero. This can be realized by our lattice model
if we set $u_3 = \lambda =0$ and
$u_1 = u_2 \ne 0$. This corresponds to wire 3 being disconnected from wires 1 
and 2; the latter two have perfect transmission into each other. Wire 3
can be bosonized as in case I, while wires 2 and 3 can be bosonized as a 
single infinite wire. Similarly, there are two other cases, called cases III
and IV, which are obtained from case II by cyclically permuting the three 
wires. We note that cases I - IV are all invariant under time reversal, if we 
choose all the entries of the $S$-matrix to be real.
 
\noindent (c) Cases V-VI: In case V, $|t_{21}|=|t_{32}|=|t_{13}|=1$, and 
all the other entries of $S$ are zero. 
No matter how the phases of the three non-zero entries
of $S$ are chosen, this is not invariant under time reversal. (Therefore it 
cannot be realized by our lattice model for any choice of the parameters 
$u_i, \lambda$ and $k_F$). This can be thought of as three infinite wires with
chiral fields; for instance, one such wire is the incoming field along wire 1 
which transmits perfectly into the outgoing field along wire 2. Finally, we 
have case VI obtained by time reversing case V; namely, $|t_{12}|=|t_{23}|
=|t_{31}|=1$, and all the other entries of $S$ are zero. Cases V and VI can 
both be bosonized.

Before ending this section, we would like to make some remarks about the
physical applicability of cases V and VI. If we think of the three wires
as having finite widths, with the incoming and outgoing waves running
along two different edges of each wire, then the forms of the $S$-matrices
in cases V and VI are very similar to those describing the edge states of
a quantum Hall system. However, the value of $K$ in a quantum Hall system
is fixed by the filling fraction of the (two-dimensional) bulk of the system, 
not by the interaction between the edge states. (In fact, the interactions
between the states at the opposite edges of a quantum Hall system are often 
ignored because of their spatial separation). In contrast to this, our model 
of the Tomonaga-Luttinger liquids in the wires and our derivation of the RG 
equations for the $S$-matrix given below both depend on the short-range 
interaction between the incoming and outgoing modes on each wire. Hence the 
results obtained by us may not be applicable to quantum Hall systems.

\section{\bf Renormalization Group Equations for the $S$-matrix}

Rather than employ bosonization to study the case of an arbitrary $S$-matrix, 
we use an instructive and physically transparent method introduced in 
Ref. \cite{yue} to directly obtain RG equations for the entries of the 
$S$-matrix in the presence of electron interactions (provided that the 
interactions are weak). The basic idea of this method is the following.
In the presence of a non-zero reflection amplitude $r_{ii}$, the density of 
noninteracting fermions in wire $i$ has Friedel oscillations with 
wavenumber $2k_F$. When a weak 
interaction is turned on, an electron scatters from these oscillations by an 
amount proportional to the parameter $\alpha_i$. Yue et al use this idea to 
derive the RG equations for an arbitrary $S$-matrix located at the junction 
of two semi-infinite wires. In the limits of both weak scattering ($r_{11} 
\rightarrow 0$) and strong scattering ($|r_{11}| \rightarrow 1$), their results
reduce to those known from bosonization \cite{kane,gogolin}. We will use the 
same idea for a junction of more than two wires. Not surprisingly, we will 
find that the results are much richer than those for two wires.

Let us briefly present the method of Yue et al. We first derive the form of
the density oscillations in one particular wire given that there is a
reflection coefficient $r$ for waves coming in along that wire. For a 
momentum in the vicinity of $k_F$, we can write the wave function in the form
\beq
\psi_k (x) ~=~ e^{-i(k+k_F)x} ~+~ r ~e^{i(k+k_F)x} ~,
\eeq
where $|k| << k_F$. In the ground state of the noninteracting system, 
the density is given by
\beq
< \rho (x) > ~=~ \int_{-\infty}^0 ~\frac{dk}{2\pi} ~\psi_k^\star (x) 
\psi_k (x) ~,
\eeq
where we have used the fact that only states with energy less than $E_F$
(i.e., momenta less than $k_F$) are occupied, and we have extended the lower 
limit to $-\infty$ for 
convenience. (Alternatively, we can impose a cut-off at the lower limit
of the form $\exp (\epsilon k)$, and take the limit $\epsilon \rightarrow
0$ at the end of the calculation). We then find that $\rho$ has a constant
piece $\rho_0$ (which can be eliminated by normal ordering the density
operator), and a $x$-dependent piece given by
\beq
< \rho (x) > ~-~ \rho_0 ~=~ ~\frac{i}{4\pi x} ~(~ r^\star ~e^{-i2k_F x} - r~
e^{i2k_F x} ~) ~.
\label{fried}
\eeq
Using the expression in (\ref{dens}), we see that the expectation value 
$<\Psi_I^\dagger \Psi_I + \Psi_O^\dagger \Psi_O>$ is a constant, while
\bea
<\Psi_O^\dagger \Psi_I> ~&=&~ ~\frac{ir^\star}{4\pi x} ~, \nonu \\
<\Psi_I^\dagger \Psi_O> ~&=&~ - ~\frac{ir}{4\pi x} ~.
\label{expec}
\eea
Note that there is also a contribution to $\rho (x)$ from the waves transmitted
from the other wires; however those are independent of $x$ and can be absorbed
in $\rho_0$. Thus the Friedel oscillations Eq. (\ref{fried}) in a given wire
only arise from reflections within that wire. 

Next we derive the reflection of the fermions from the Friedel oscillations,
using a Hartree-Fock decomposition of the interaction in Eq. (\ref{hint2}).
The reflection is caused by the following terms in the decomposition
\bea
H_{\rm int} ~&=&~ - g_2 ~\int_0^\infty ~dx ~(~ <\Psi_O^\dagger \Psi_I> 
\Psi_I^\dagger \Psi_O \nonu \\
& & ~~~~~~~~~~~~~~~~~~~~~~+~ <\Psi_I^\dagger \Psi_O> \Psi_O^\dagger \Psi_I ~)~,
\nonu \\
&=&~ - \frac{ig_2}{4\pi} ~\int_0^\infty ~\frac{dx}{x} ~( r^\star ~
\Psi_I^\dagger \Psi_O ~-~ r ~\Psi_O^\dagger \Psi_I )~,
\label{hf1}
\eea
where we have used (\ref{expec}) to write the second equation. Now we can
derive the amplitude to go from a given incoming wave with momentum $k$
to an outgoing wave (or vice versa) under the action of $\exp (-iH_{\rm int}
t)$. The amplitude is given by
\bea
& & - i ~\int \frac{dk^\prime}{2\pi} ~2\pi \delta (E_k - E_{k^\prime}) ~
|{\rm outgoing}, k^\prime > \nonu \\
& & ~~~~~~~~~~~~~~~~~~~~~~~~ \times <{\rm outgoing}, k^\prime | ~H_{\rm 
int} ~| {\rm incoming}, k> \nonu \\
& &~ =~ |{\rm outgoing}, k> ~\frac{ig_2 r}{4\pi \hbar v_F} ~\int_0^\infty ~
\frac{dx}{x}~ e^{-i2k x} ~, 
\label{trans}
\eea
where we have used Eq. (\ref{hf1}), the dispersion relation $E_k = \hbar v_F 
k$ (so that $\delta (E_k - E_{k^\prime}) = (1/\hbar v_F ) \delta (k - 
k^\prime)$), and the wave
functions $\exp (\pm ikx)$ of the outgoing and incoming waves respectively.
The integral over $x$ in (\ref{trans}) is divergent at the lower end;
we therefore introduce a short-distance cut-off $d$ there. The amplitude
in (\ref{trans}) then reduces to 
\beq
- \frac{\alpha r}{2} ~{\rm ln} (kd)
\label{inout}
\eeq
plus pieces which remain finite as $kd \rightarrow 0$; we have used Eq. 
(\ref{ali}) here. Similarly, the amplitude to go from an outgoing wave to an 
incoming wave is given by
\beq
\frac{\alpha r^\star}{2} ~{\rm ln} (kd) ~.
\label{outin}
\eeq

These reflections from the Friedel oscillations can then be combined along 
with the $S$-matrix at the junction to calculate the corrections to the 
$S$-matrix. For instance, consider $r_{ii}$. To first order in the interaction
parameters $\alpha_i$, this gets corrections from the following processes. 
An incoming wave on wire $i$ can either (i) turn into an outgoing wire
on the same wire with the amplitude in (\ref{inout}) (with $r$ replaced
by $r_{ii}$ in that expression), or (ii) get reflected 
from the junction with amplitude $r_{ii}$ thereby turning into an outgoing 
wave, turn back into an incoming wave according to (\ref{outin}), then get 
reflected again from the junction, or (iii) transmit through the junction into
wire $j$ (with $j \ne i$) with amplitude $t_{ji}$, turn from an outgoing wave 
to an incoming wave on wire $j$ according to (\ref{outin}) (with $r$ replaced 
by $r_{jj}$), then transmit back through the junction to wire $i$ 
with amplitude $t_{ij}$. The correction to $r_{ii}$ is therefore
\bea
dr_{ii} ~&=&~ - ~A_{ii} {\rm ln} (kd) ~, \nonu \\
{\rm where} \quad A_{ii} ~&=&~ -~ \frac{1}{2} ~[~ - ~\alpha_i r_{ii} ~
+~ \alpha_i |r_{ii}|^2 r_{ii} \nonu \\
& & ~~~~~~~~~~~ +~ \sum_{j \ne i} ~\alpha_j t_{ij} r_{jj}^\star t_{ji} ~]~.
\label{aii}
\eea
Similarly, the transmission amplitude $t_{ji}$ from 
wire $i$ to wire $j$ can get corrections from the following processes.
The incoming wave on wire $i$ can either (i) get reflected from the junction
with amplitude $r_{ii}$, turn back into an incoming wave according to 
(\ref{outin}), and then transmit into wire $j$ with amplitude $t_{ji}$,
or (ii) transmit into to wire $j$ first, turn into an incoming wave on wire 
$j$ according to (\ref{outin}), then get reflected from the junction with 
amplitude $r_{jj}$, or (iii) transmit into a wire $k$ (with $k \ne i,j$), turn
into an incoming wave in wire $k$ according to (\ref{outin}) (with $r$ 
replaced by $r_{kk}$), then transmit into wire $j$ with amplitude $t_{jk}$.
Hence the correction to $t_{ji}$ is 
\bea
dt_{ji} ~&=&~ - ~A_{ji} {\rm ln} (kd) ~, \nonu \\
{\rm where} \quad A_{ji} ~&=&~ - ~\frac{1}{2} ~[~ \alpha_i t_{ji} |r_{ii}|^2 ~
+~ \alpha_j |r_{jj}|^2 t_{ji} \nonu \\
& & ~~~~~~~~~~ +~ \sum_{k \ne i,j} ~\alpha_k t_{jk} r_{kk}^\star t_{ki} ~]~.
\label{aji}
\eea

Yue et al then derive the RG equations for the $S$-matrix which is now 
considered to be a function of a distance scale $L$; they show that $-{\rm 
ln} (kd)$ in Eqs. (\ref{aii}-\ref{aji}) can effectively be replaced by $dl$, 
where $l = {\rm ln} (L/d)$. The RG equations therefore take the from
\bea
\frac{dr_{ii}}{dl} ~&=&~ A_{ii} ~, \nonu \\
\frac{dt_{ij}}{dl} ~&=&~ A_{ij} ~,
\label{rg1}
\eea
where $A_{ii}$ and $A_{ij}$ are given above. We can write Eqs. (\ref{rg1}) in 
a simpler way. Given the matrix $S$ and the parameters $\alpha_i$ (which do
not flow under RG), we can define a diagonal matrix $F$ whose entries are 
\beq
F_{ii} ~=~ - ~\frac{1}{2} ~\alpha_i r_{ii} ~.
\label{fmat}
\eeq
Then the RG equations can be written in the matrix form
\beq
\frac{dS}{dl} ~=~ S F^\dagger S ~-~ F ~.
\label{rg2}
\eeq
This is the central result of our work. One can verify from (\ref{rg2}) that 
$S$ continues to remain unitary under the RG flow; it also remains symmetric 
if it begins with a symmetric form. 

We note also that the form of (\ref{rg2})
remains unchanged if $S$ is multiplied either from the left or from the right 
by a diagonal unitary matrix with entries of the form 
\beq
U_{ii} ~=~ e^{i\phi_i} ~,
\label{phase}
\eeq
where the real numbers $\phi_i$ are independent of the length parameter $l$.
The fixed points discussed below will therefore also remain unchanged under
such phase transformations. We will generally not distinguish between 
$S$-matrices which differ only by such transformations.

\section{\bf Fixed Points And Stability Analysis}

We will now study the RG flow in some detail. We will consider the case
of three wires for convenience, although much of the discussion below can be
generalized to more than three wires. Let us first find the fixed points
of Eq. (\ref{rg2}). The required condition is that $S F^\dagger = F 
S^\dagger$, i.e., that $S F^\dagger$ is hermitian. It is easy to see that
the six cases I - VI considered above are all fixed points of the RG. In
addition, there is another fixed point which we will call case VII. For the
physically interesting situation in which all the $\alpha_i$ are positive,
this case is described as follows. We first define a quantity $a$ as
\beq
a ~=~ \frac{1}{\sum_{i=1}^3 ~\alpha_i^{-1}} ~.
\eeq
Then the fixed point $S$-matrix has the entries
\bea
r_{ii} ~&=&~ - ~\frac{a}{\alpha_i} \quad {\rm for ~all ~} i ~, \nonu \\
t_{ij} ~&=&~ ~\sqrt{(1 ~-~ \frac{a}{\alpha_i}) (1 ~-~ \frac{a}{\alpha_j})}
\quad {\rm for ~all ~} i,j ~.
\eea
It is possible to obtain a family of fixed points related to the above by
multiplying the various amplitudes by some phases as discussed in Eq. 
(\ref{phase}). However, we will mainly
consider the above form of case VII for simplicity. Note that for the case
of equal interactions in the three wires (i.e., all the $\alpha_i$ equal to
each other), and $\lambda =0$, the fixed point is a well-known $S$-matrix whose
entries are $r_{ii} = -1/3$ for all $i$, and $t_{ij} =2/3$ for all $i,j$.
This is symmetric under all possible permutations of the three wires, and has
the maximum transmission (in all channels simultaneously) allowed by unitarity.

Having found the fixed points of the RG equations, we can study their
stabilities. Let us write a fixed point of the $S$-matrix as $S_0$, and a 
small deviation from this as the matrix $\epsilon S_1$, where $\epsilon$ is 
a small real parameter. Given $S_0$, we are interested in finding the various
flow `directions' $S_1$ such that Eq. (\ref{rg2}) takes the simple form 
\beq
\frac{d\epsilon}{dl} ~=~ \mu \epsilon ~,
\label{dedl}
\eeq
where $\mu$ is a real number.
The solution of this equation is $\epsilon (l) = \epsilon (0) \exp (\mu 
l)$, where $\epsilon (0)$ is given by the deviation of the $S$-matrix from
$S_0$ at the microscopic (e.g., lattice) scale. Thus $\mu < 0$ indicates that
$S$ is stable against a perturbation in the direction of the corresponding
$S_1$, while $\mu > 0$ indicates an instability in the direction of the
corresponding $S_1$. We now consider the various fixed points in turn. [All 
the fixed points have directions in which $\mu =0$ corresponding to the phase 
rotations of the $S$-matrix described in Eq. (\ref{phase}). We will ignore 
these directions in the following discussion.] 

\noindent (a) Case I: This turns out to be stable against perturbations in all
directions. There are three directions in which $\mu$ takes the values 
$-(\alpha_1 + \alpha_2)/2$, $-(\alpha_2 + \alpha_3)/2$, and $-(\alpha_1 + 
\alpha_3)/2$ respectively. These are negative since we are assuming that the 
interactions in all the wires are repulsive. Note that this result agrees, to 
first order in the $\alpha_i$, with the exact results one obtains from 
bosonization. The operator which tunnels a particle from wire 1 to wire 2 has 
the scaling dimension $(K_1 + K_2)/(2 K_1 K_2)$. For weak interactions, this 
is equal to $1 + (\alpha_1 + \alpha_2)/2$. Under a RG flow, therefore,
the coefficient of the tunneling operator satisfies Eq. (\ref{dedl}) with
$\mu = - (\alpha_1 + \alpha_2)/2$.

\noindent (b) Cases II-IV: Case II has two stable directions, both with 
$\mu = - 
\alpha_3 /2$ (these correspond to tunneling from wire 3 to wire 1 or wire 2),
and one unstable direction with $\mu = (\alpha_1 + \alpha_2)/2$ corresponding
to reflection between wires 1 and 2. These results also agree, to first
order in the $\alpha_i$, with those obtained from bosonization. For tunneling
from wire 3 to wire 1, the operator has the dimension $1/(2K_3) + (K_1 + 
1/K_1)/4$; this is equal to $1 + \alpha_3 /2$ to first order in the $\alpha_i$,
and therefore gives $\mu = -\alpha_3 /2$. A weak reflection between wires
1 and 2 has the dimension $(K_1 + K_2)/2$ which is equal to $1 - (\alpha_1 +
\alpha_2)/2$ to first order. This gives a flow with $\mu = (\alpha_1 + 
\alpha_2)/2$ which goes in the direction of case I. The RG flows in cases 
III and IV can be worked out similarly.

\noindent (c) Cases V-VI: Case V has three unstable directions with $\mu = 
\alpha_1/2$, $\alpha_2/2$ and $\alpha_3/2$ respectively. The three directions
give flows towards cases II, III and IV. However, if we start out with an 
appropriate linear combination of the three directions, we can flow directly 
to case I. Similarly, one can work out the flows for case VI. 

\noindent (d) Case VII: The stability of flows near this fixed point appears
to be difficult to study in general. However, the case of equal interactions,
$\alpha_1 = \alpha_2 = \alpha_3 = \alpha$, can be studied more easily. It has
one unstable direction with $\mu = \alpha$ which flows in the direction of
case I (this is discussed further in Eq. (\ref{flow71}) below), and two 
unstable directions with $\mu = 2\alpha /3$ which flow towards one of the
four cases I - IV depending on the precise choice of the initial direction. 
Further, for appropriately chosen directions of the initial flow, one can go 
from cases V and VI to case VII. Near case VII, these correspond to two 
stable directions with $\mu = - \alpha$. 

\begin{figure}[htb]
\begin{center}
\epsfig{figure=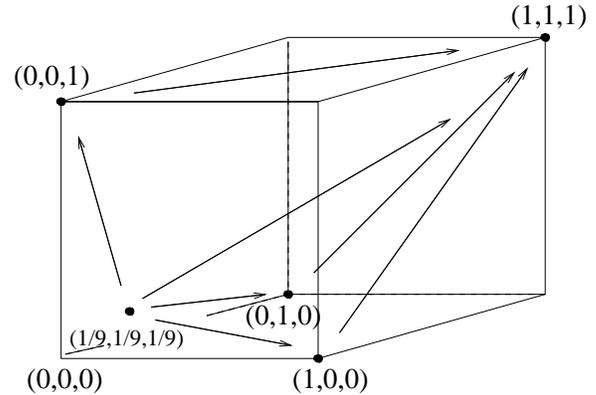,width=8cm}
\end{center}
\caption{Schematic diagram of the various time-reversal invariant fixed 
points for the 3-wire junction problem. Cases I-IV and VII are indicated as 
(1,1,1), (0,0,1), (0,1,0), (1,0,0) and (1/9,1/9,1/9), where the sets of 
three numbers denote the modulus squared of the diagonal entries of the 
respective $S$-matrices. RG flows between the various fixed points are 
indicated by the arrows.}
\end{figure}

Based on the above, we can state the flow diagram in the space of all 
$S$-matrices as follows. In general, case I is the most stable. Cases II, III 
and IV are only unstable to a flow towards case I. Cases V and VI are unstable
to flows towards cases I - IV. Finally, for the case of equal interactions
$\alpha_i$, case VII is unstable to flow towards cases I - IV, and cases
V and VI are also unstable to a flow towards case VII (if one starts out
in the appropriate direction). We have verified this flow diagram numerically
by starting from a number of $S$-matrices close to the various fixed points 
and letting them evolve according to Eq. (\ref{rg2}). 

We thus see that the flow diagram for the case of 3 wires (with repulsive 
interactions on all the wires) is much richer than in the case of two wires. 
In the latter case, there are only two fixed points, a stable one at $|r_{11}| 
= |r_{22}| = 1$ (two disconnected wires), and an unstable one at $|t_{12}|
= |t_{21}| = 1$ (a perfectly transmitting wire). The RG flow simply goes from
the first point to the second \cite{kane,gogolin}. 

\section{\bf Conductance Of A Three-Wire System}

Having studied the $S$-matrix for a three-wire system as a function of the
RG distance scale $l$, we can now discuss the conductance of this system.
We will assume that the three wires, instead of being really semi-infinite,
are connected to three Fermi liquid leads (with the interaction parameter 
being given by $K=1$) at a large distance from the junction. We will also
assume that there is only one transverse channel of spinless fermions in 
each wire; in this band, there is a resistance of $e^2 /h$ at the contacts 
between the leads and the wires \cite{buttiker}. Although the contacts can 
themselves scatter the fermions \cite{lal}, we will ignore such effects here. 

We take the fermions in all the leads to have the same Fermi energy $E_F$, 
and the net current on all wires to be zero in the absence of any applied 
voltage on the leads. Now suppose that the voltage is changed by a small 
amount $V_i$ on lead $i$. Then the net current flowing out of wire $i$ will 
satisfy the linear relationship \cite{buttiker1,buttiker}
\beq
I_i ~=~ \frac{e^2}{h} ~\sum_{j=1}^3 ~T_{ij} V_j ~,
\label{cond1}
\eeq
where the $T_{ij}$ (for $i \ne j$) define the various transmission 
probabilities, and $T_{ii} +1$ denote the reflection probabilities. These are
related to the $S$-matrix at the junction as follows
\bea
T_{ij} ~&=&~ |t_{ij}|^2 \quad {\rm for} \quad i \ne j ~, \nonu \\
{\rm and} \quad T_{ii} ~&=&~ |r_{ii}|^2 ~-~ 1 ~. 
\label{cond2}
\eea
Since the unitarity of the $S$-matrix implies that 
\beq
|r_{ii}|^2 ~+~ \sum_{j \ne i} ~|t_{ij}|^2 ~=~ 1 ~,
\label{unitarity}
\eeq
we see from (\ref{cond1}) that the currents $I_j$ do not change if all the
voltages $V_i$ are changed by the same amount. When a small voltage 
$V_i$ is applied on lead $i$ in addition to the Fermi energy, it increases the
number of incoming fermions on that lead by an amount given by $eV_i$ times the
density of states in energy per unit length. For noninteracting spinless
fermions in one dimension, the density of states in a continuum theory is 
given by
\beq
\rho (E_F) ~=~ \frac{1}{2\pi \hbar v_F} ~,
\eeq
where $v_F$ is the Fermi velocity. We assume this expression for $\rho (E_F)$
to be the same on all leads. In the absence of any scattering from the 
contact $i$ or from impurities inside wire $i$, these fermions will travel 
ballistically towards the junction where they
will be either reflected back or transmitted to one of the other two wires. 
Following that, the fermions again travel ballistically till they emerge
from one of the three wires. The outgoing currents are therefore given by 
$ev_F$ times the extra number of electrons coming in on wire $i$ times the 
appropriate transmission coefficients on the other two wires and the 
reflection coefficient (subtracted from the incoming current) on wire $i$. 

We can now compute the conductance by setting, say, wire 3 to be the 
potential probe, i.e., $I_3 = 0$ \cite{buttiker1}. Then, using the 
set of equations (\ref{cond1})-(\ref{unitarity}) given above, the corresponding
three-terminal relations are found to be
\bea
G_{12,13} &=& \frac{I_1}{V_1 -V_3} = \frac{e^2}{h} ~(T_{12}+T_{13}
+\frac{T_{12}T_{13}}{T_{32}}) ~, \\
G_{12,23} &=& \frac{I_1}{V_2 -V_3} = \frac{e^2}{h} ~(T_{12}+T_{32}
+\frac{T_{12}T_{32}}{T_{13}}) ~,
\label{cond3}
\eea 
where $I_1 = - I_2$, and the two-terminal conductance is given by
\beq
G_{12,12} = \frac{I_1}{V_1 -V_2} = \frac{e^2}{h} ~(T_{12} + 
\frac{T_{13}T_{32}}{T_{13}+T_{32}}) ~.
\label{cond4}
\eeq
In the above conductance expressions, we have employed the standard 
convention for specifying the current (first pair of indices) and voltage 
(second pair of indices) leads. It is worth noting the incoherence introduced 
in $G_{12,12}$ through the non-zero transmissions of carriers $T_{13}$ and 
$T_{32}$ into the additional arm (here, wire 3). The conductances given in 
Eqs. (\ref{cond3}) and (\ref{cond4}) will flow under RG following Eq. 
(\ref{rg2}). Let us begin with some $S$-matrix at a microscopic 
distance scale $d$ (such as the spacing between the sites in a lattice 
model). The RG flow in (\ref{rg2}) is valid till the logarithmic
length scale reaches a physical long-distance cut-off. The appropriate 
cut-off in this problem is the smaller of the scales ${\rm ln} (L_i /d)$ 
(where $L_i$ is the length of wire $i$ from the junction to its lead) and 
${\rm ln} (L_T /d)$, where
\beq
L_T ~=~ \frac{\hbar v_F}{k_B T} ~,
\eeq
with $T$ being the temperature \cite{lal}. For simplicity, let us consider 
the case of high temperature where $L_T$ is smaller than all the wire lengths
$L_i$, but larger than the microscopic length $d$. Then the RG flow has to be 
stopped at the scale $l_T = {\rm ln} (L_T /d)$ since there is no phase 
coherence at distance scales larger than this. Now let us suppose that at the 
microscopic level, the $S$-matrix deviates slightly from a fixed point $S_0$ 
by an amount $\epsilon (0) S_1$, where $S_1$ is an unstable direction with 
$\mu > 0$. Then at the scale $l_T$, the deviation is given by 
\beq
dS (l_T) ~=~ \bigl( \frac{L_T}{d} \bigr)^\mu \epsilon (0) S_1 ~.
\label{dev}
\eeq
We thus see that the deviations from $S_0$ will grow as $1/T^\mu$ as the 
temperature decreases. Of course, this is only true as long as the deviation
is not too large, since Eq. (\ref{dedl}) is only valid to first order in 
$\epsilon$. These power-law dependences of the conductances on the temperature
should be observable experimentally if a three-wire system can be fabricated.

As a specific example, consider case VII in which $S_0$ has $r_{ii} =-1/3$ and
$t_{ij} =2/3$. If all the interactions are equal, with $\alpha_i = \alpha$, we
saw above that this is unstable to a perturbation towards case I (three
disconnected wires) with $\mu = \alpha$. The small deviation which takes 
case VII towards case I is given by
\bea
dS ~=~ \left( \begin{array}{ccc} i 4\epsilon - 3\epsilon^2 & i \epsilon - 15 
\epsilon^2 /2 & i \epsilon - 15 \epsilon^2 /2 \\
i \epsilon - 15 \epsilon^2 /2 & i 4\epsilon - 3\epsilon^2 & i \epsilon - 15 
\epsilon^2 /2 \\ 
i \epsilon - 15 \epsilon^2 /2 & i \epsilon - 15 \epsilon^2 /2 & i 4\epsilon - 
3\epsilon^2 \end{array} \right) 
\label{flow71}
\eea
to second order in the real parameter $\epsilon$. We have gone up to second
order so as to calculate the correction to $T_{ij}$ which only begins at
that order. Namely, $t_{ij} = 2/3 + i \epsilon - 15 \epsilon^2 /2$, which 
gives $T_{ij} = 4/9 - 9 \epsilon^2$. Since $\mu = \alpha$, the 
deviation of $T_{ij}$ from $4/9$ will grow as $\epsilon^2 \sim 
1/T^{2\alpha}$ as the temperature is reduced. 
For example, the two-terminal conductance in this case will be 
\beq
G_{12,12}^{VII} = \frac{e^2}{h}(\frac{2}{3} - \frac{27}{2} c_1 T^{-2\alpha}) ~,
\eeq
where $c_1$ is some constant, while the three-terminal conductances for this 
case are identical and are given by
\beq
G_{12,13}^{VII} = \frac{e^2}{h} (\frac{4}{3} - 27 c_1 T^{-2\alpha}) 
\simeq 2 G_{12,12}^{VII}
\eeq
Thus the power-law dependence on $T$ can provide information on the 
strength of the interaction $\alpha$. 

\section{\bf Four-Terminal Conductance of a Quantum Wire}

We consider here the case of the four-terminal conductance of a quantum 
wire of finite length $L$ measured at high temperatures such that 
the thermal length $L_T ~(=\hbar v_F /k_B T)~> l$. 

\begin{figure}[htb]
\begin{center}
\epsfig{figure=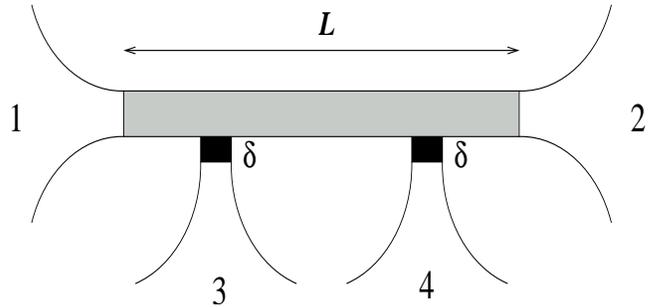,height=4cm,width=8.5cm}
\end{center}
\caption{Schematic diagram of a quantum wire of length $L$ 
(grey shaded region) connected to the two current probes ($1$ and $2$) 
and two voltage probes ($3$ and $4$). The voltage probes are very 
weakly coupled to the quantum wire via tunneling barriers of
amplitude $\delta \ll 1$ (black shaded regions).}
\end{figure}

The quantum wire is connected to two reservoirs $1$ and $2$ which act as 
current probes. In addition, the quantum wire is also weakly coupled to two 
voltage probes $3$ and $4$ via identical barriers with tunneling amplitudes 
$\delta \ll 1$. We consider the current and voltage probes to be 
semi-infinite, two-dimensional electron gas (2DEG) Fermi liquid reservoirs; 
these can be modeled as Tomonaga-Luttinger leads with interaction 
parameter $K_L =1$, i.e., in our case of very weak interactions, 
$\alpha_L =0$. The quantum wire is itself modeled as a TLL with 
weak repulsive interactions characterized by a parameter $\alpha_W$. 
We can now see that this case is akin to that of a system of 
two 3-wire junctions with a common arm (of finite length $L$ and with 
$L \ll L_T$). Further, one arm in each of the two 3-wire junctions is 
coupled to the other two through a weak tunneling amplitude; this case 
thus falls somewhere between the fixed point I and any one of the fixed 
points II-IV discussed earlier. 

Following the analysis of Ref. \cite{buttiker1}, we can write the 
four-terminal conductance of this system as
\beq
G_{12,34}=\frac{e^2}{h} T_{12}\frac{(T_{31}+T_{32})(T_{41}+T_{42})}
{T_{31}T_{42}-T_{32}T_{41}}
\label{4termcond}
\eeq
where $T_{ij}$ stands, as usual, for the transmission from lead $i$ 
to lead $j$. As transmission from lead $1$ to lead $2$ can take place 
through paths which never cross any of the two tunneling barriers, we 
can write (to lowest order in $\delta$) $T_{12}=T_{12}^{(0)}$. 
Transmission from lead $3$ to lead $1$ can take place, to lowest order 
in $\delta$, through a path that crosses one tunneling barrier; this gives 
$T_{31}=\delta T_{31}^{(1)}$, where $T_{31}^{(1)}$ is a positive number of
order 1. Similarly, even the simplest 
path from lead $3$ to lead $4$ needs the crossing of two barriers, 
giving $T_{34}=\delta^2 T_{34}^{(2)}$. Thus, keeping only terms till 
order $\delta^2$, we can write $G_{12,34}$ as 
\beq
G_{12,34}=\frac{e^2}{h} T_{12}^{(0)}\frac{(T_{31}^{(1)}+T_{32}^{(1)})
(T_{41}^{(1)}+T_{42}^{(1)})}{T_{31}^{(1)}T_{42}^{(1)}
-T_{32}^{(1)}T_{41}^{(1)}}
\label{firstcond}
\eeq
The four-terminal resistance $R_{12,34} = 1/G_{12,34}$ lies in 
the range $-h/(e^2 ~T_{12}^{(0)}) \le R_{12,34} \le h/(e^2 ~T_{12}^{(0)})$.
The RG flow of the tunneling barriers $\delta$ will take place as given 
earlier, but with a parameter $\mu$ which is dependent on the interaction
parameter of the quantum wire $\alpha_W$ (note that $\alpha_L=0$ for all the 
four probes). Now, as we have identical barriers 
connecting two identical voltage probes $3$ and $4$ to the quantum wire, the 
various wire-voltage probe transmissions, such as $T_{31}^{(1)}, T_{32}^{(1)},
T_{41}^{(1)}$ and $T_{42}^{(1)}$, will have identical power-law dependences 
on the temperature. Thus, in the expression (\ref{firstcond}) given above, 
the only temperature dependence of $G_{12,34}$ will come from 
the RG flow of the transmission $T_{12}^{(0)}$, since all the temperature 
dependences coming from the wire-voltage probe transmissions will cancel out. 
Further, as the two-terminal conductance $G_{12,12} 
\sim (e^2 /h)~T_{12}^{(0)}$ to lowest order in $\delta$, we can see 
that the temperature dependences of $G_{12,12}$ and $G_{12,34}$ are 
identical! 


In a recent experiment \cite{picci}, Picciotto {\it et al.} measured 
both the two-terminal resistance $R_{12,12}~(=1/G_{12,12})$ as well as the 
four-terminal resistance $R_{12,34}$ of a cleaved-edge overgrowth 
quantum wire in a GaAs-AlGaAs heterojunction using two weakly 
coupled voltage probes. They found that while 
$R_{12,12}$ is quantized in integer multiples of $h/2e^2$, $R_{12,34}$ 
fluctuated above and below zero and finally vanished as the gate voltage 
was made less negative. $R_{12,34}$ fluctuates about zero 
because its value depends critically
on the invasive nature of the probes (i.e., on the precise values of the
transmissions $T_{31}^{(1)}, T_{32}^{(1)}, T_{41}^{(1)}$ and $T_{42}^{(1)}$), 
and the fact that $-h/(e^2 T_{12}^{(0)}) \le R_{12,34} \le 
h/(e^2 T_{12}^{(0)})$. The average value of $R_{12,34}$ vanishes due to 
the fact that the intrinsic resistance of a quantum wire without any 
defects or impurities comes from its connections to the 2DEG reservoirs, 
i.e., the so-called {\it contact resistances} \cite{picci}. Our prediction of 
the identical power-law variations of $G_{12,12}$ and $G_{12,34}$ with 
temperature can also be tested in such an experiment by taking 
measurements of the two conductances at various temperatures but at a 
fixed value of the gate voltage (this holds the values of the various 
transmissions $T_{ij}$ fixed at the microscopic level, and their
observed values will vary with the temperature through the RG equations).

\section{\bf Analysis Of A Four-Wire System}

We can carry out a similar analysis of the fixed points and the conductance 
for a system of four wires meeting at a junction. In this section, we will 
assume for simplicity that the interaction parameters $\alpha_i =\alpha$ 
are equal on all the wires. 

Let us first consider the fixed points of the RG equations (\ref{rg2}) for a 
four-wire system. To begin with, one can readily identify $4! =24$ fixed 
points which are natural generalizations of the $3! =6$ fixed points 
(cases I-VI) that we found above for the three-wire system. These fixed 
points correspond to all the possible ways in which each row (or column) of 
the $S$-matrix at the junction has only one non-zero entry whose modulus is 
equal to 1. We thus have the following possibilities. 

\noindent (a) The simplest case is one in which all the four wires are 
disconnected from each other. The $S$-matrix is then diagonal, with all the 
diagonal entries having unit modulus.

\noindent (b) There are six cases in which two of the wires are disconnected
from all the others, while the remaining two wires transmit perfectly into
each other.

\noindent (c) There are three cases in which pairs of wires (say, 1,2 and 
3,4) transmit perfectly into each other.

\noindent (d) There are eight cases in which one wire (say, wire 4) is 
disconnected from the other three, while the other three wires (1, 2 and 3)
are connected to each other cyclically as in Cases V and VI for the 
three-wire system.

\noindent (e) There are six cases in which the four wires transmit perfectly
into each other in a cyclical way, such as, 1 into 2, 2 into 3, 3 into 4, 
and 4 into 1.

We note that the 10 cases given in (a-c) are invariant under time reversal
if we choose all the entries of the $S$-matrix to be real; these 10 cases
allow bosonization to be done. The 14 cases in (d-e) necessarily violate 
time-reversal invariance; they can also be bosonized.

In addition to the 24 cases given above, there are 5 more fixed points of the
RG equations. Four of these correspond to situations in which one of the
wires (say, 4) is disconnected from the other three, while the other
three wires (1, 2 and 3) have the completely symmetric and maximally 
transmitting $S$-matrix of the form given in Case VII above. The fifth case
corresponds to the case in which the four wires have a completely
symmetric and maximally transmitting $S$-matrix; the diagonal and off-diagonal
entires of this matrix are given by $-1/2$ and $1/2$ respectively.

We thus have a total of 29 fixed points for a four-wire system in contrast to
7 fixed points for the three-wire system. In addition to these 29 cases, we 
will now see that the four-wire system has some new classes of fixed points 
which do not exist for systems with less than four wires. Namely, there exist
{\it two-parameter families} of fixed points in the four-wire system. In
contrast to these, the fixed points of the two- and three-wire systems are
all isolated points, i.e., they have no variable parameters (apart from some 
trivial phases). 

Although we have not studied all the two-parameter families of
fixed points in the four-wire system, we can exhibit some of these families
explicitly. Two examples are given by
\bea
S ~=~ \left( \begin{array}{cccc} 0 & x_1 & iy_1 & 0 \\
x_2 & 0 & 0 & iy_2 \\
iy_2 & 0 & 0 & x_2 \\ 
0 & iy_1 & x_1 & 0 \end{array} \right) ~,
\label{smat1}
\eea
where $x_i$ and $y_i$ are four real numbers satisfying the constraints 
$x_1^2 + y_1^2 = x_2^2 + y_2^2 =1$, and
\bea
S ~=~ \left( \begin{array}{cccc} 0 & x & -y & -z \\
x & 0 & -z & y \\
-y & z & 0 & x \\ 
z & y & x & 0 \end{array} \right) ~,
\label{smat2}
\eea
where $x, y$ and $z$ are three real numbers satisfying the constraint $x^2
+y^2 + z^2 =1$. [It is easy to see that these are fixed points of Eq. 
(\ref{rg2}) since the diagonal matrix $F$ is equal to zero for these families].
Note that these two families have some members in common
which are obtained by setting $x_1 = x_2 =x$, $y_1 = y_2 =y$ and $z=0$, and
then performing some phase transformations. Further, these families include 
some of the fixed points given earlier as special cases. 

The two-parameter families are fixed points of the RG equations (\ref{rg2}) 
which are only valid to first order in the interaction parameter $\alpha$.
Do they remain fixed points if we go to higher orders in $\alpha$?
One way to answer this question is to use the technique of bosonization.
As remarked earlier, it does not seem possible to bosonize an interacting
fermionic theory for all possible $S$-matrices. Fortunately, 
the two-parameter families described above contain some special points at 
which bosonization can be done. For instance, consider the $S$-matrix
\bea
S ~=~ \left( \begin{array}{cccc} 0 & 1 & 0 & 0 \\
1 & 0 & 0 & 0 \\
0 & 0 & 0 & 1 \\ 
0 & 0 & 1 & 0 \end{array} \right) ~,
\label{smat3}
\eea
which corresponds to wires 1,2 (and wires 3,4) transmitting perfectly into 
each other. We can bosonize this system; for equal interaction strengths on 
all the wires, the bosonic theory will have the same parameter $K$ for all 
wires. We then turn on small perturbations corresponding to either $y_1, y_2$ 
in the 
family given in (\ref{smat1}), or $y, z$ in the family given in (\ref{smat2}).
These correspond to hopping at the junction between wire 1 (or 2) and wire
3 (or 4). All these hopping operators have the scaling dimension $(K + 1/K)/2$
which is necessarily larger than 1; hence they are irrelevant, and the
perturbed $S$-matrices will therefore flow back to Eq. (\ref{smat3}) under RG. 
For weak interactions with $K=1-\alpha$, we see that the scaling dimension 
differs from 1 only at order $\alpha^2$ and higher, which explains why these 
small perturbations look like fixed points at order $\alpha$.

We therefore conclude that the two-parameter families given above are 
generally {\it not} fixed points of the exact (i.e., to all orders in the 
interaction strengths) RG equations. Although we have shown this only in the 
vicinity of some bosonizable points, it is plausible that this statement will 
also be true for most other members of the families. However, this does not
rule out the possibility that there may be non-trivial and isolated 
members of these families which are fixed points of the exact RG equations.
Let us present a plausible example of such a non-trivial fixed point.
We consider a one-parameter family of $S$-matrices of the form
\bea
S ~=~ \left( \begin{array}{cccc} 0 & x & iy & 0 \\
x & 0 & 0 & iy \\
iy & 0 & 0 & x \\ 
0 & iy & x & 0 \end{array} \right) ~,
\eea
where $x^2 + y^2 =1$ and $0 \le x,y \le 1$. The two end-points of this
family given by $(x,y)=(1,0)$ and $(0,1)$ are bosonizable because they consist
of pairs of perfectly transmitting wires (1,2 and 3,4 at the first point, 
and 1,3 and 2,4 at the
second point) which transmit perfectly into each other. Within this
one-parameter family, the bosonization approach discussed above shows that
both the end-points are stable, since small perturbations from them 
(corresponding to turning on $y_1 = y_2$ in Eq. (\ref{smat1})) are
irrelevant. The simplest possibility therefore is that there is one
unstable fixed point which lies between the two end-points; since the
interaction strengths in all the wires are equal, this fixed point is likely
to be at the half-way point given by $x=y=1/{\sqrt 2}$. However, we are 
unable to directly verify that this is an unstable fixed point of the exact
RG equations since this point does not seem to be bosonizable.

To summarize, we see that the pattern of fixed points and RG flows for a 
four-wire system is immensely more complicated than those of two- and 
three-wire systems. We do not have a complete classification of the fixed 
points for a four-wire system. Some families of $S$-matrices which appear 
to be fixed points at first order in the interaction strengths turn
out not to be fixed points at higher orders.

We now turn to a discussion of the temperature dependences of the conductance
corrections. Our arguments will be very similar to those given for a three-wire
system at the end of Sec. VI. We consider the vicinity of one particular 
fixed point of the four-wire system, namely, the completely symmetric and 
maximally transmitting $S$-matrix. Let us perturb this in a completely 
symmetric way, so that the entries of the $S$-matrix are given by 
\bea
r_{ii} ~&=&~ - \frac{1}{2} ~+~ i 3 \epsilon ~-~ 3 \epsilon^2 ~, \nonu \\
t_{ij} ~&=&~ \frac{1}{2} ~+~ i \epsilon ~-~ 5 \epsilon^2 ~,
\label{randt} 
\eea
to second order in the small real number $\epsilon$. [This perturbation will 
eventually lead to the situation in which all the four wires are disconnected 
from each other]. Using Eqs. (\ref{rg2}), we find that the perturbation 
initially grows as in Eq. (\ref{dev}) with $\mu = \alpha$. The arguments 
presented in Sec. VI therefore imply that at high temperature, the 
transmission probabilities $T_{ij} = 1/4-4 \epsilon^2$ vary with 
temperature as
\beq
T_{ij} ~=~ \frac{1}{4} ~-~ c_2 ~ T^{-2\alpha} ~,
\eeq
where $c_2$ is some constant.

We can also compute the four-terminal conductances of this system 
by following the arguments of Ref. \cite{buttiker1} and those 
given in Sections VI and VII. For a set of four probes $\{m~n~k~l\}$ 
(which will be a permutation of $\{1~2~3~4\}$), we can write the relation 
between the currents $I_m =-I_n =I_1$, $I_k =-I_l =I_2$ and 
the voltages $V_1 =(\mu_m -\mu_n )/e$, $V_2 =(\mu_k -\mu_l )/e$ (where $\mu_i$
denotes the chemical potential of the $i^{\rm th}$ probe) as
\beq
\left( \begin{array}{c}
I_1 \\ I_2
\end{array} \right) ~=~ \frac{e^2}{h} ~\left( \begin{array}{c c}
\alpha_{11} & -\alpha_{12} \\ \alpha_{21} & \alpha_{22} \end{array} \right)
\left( \begin{array}{c}
V_1 \\ V_2
\end{array} \right) ~,
\eeq
where
\bea
\hspace*{-2cm}\alpha_{11} &=& \lbrack(1-T_{11})P-(T_{14} +
T_{12})(T_{41}+T_{21}) \rbrack /P ~, \nonu \\
\alpha_{12} &=& (T_{12}T_{34}-T_{14}T_{32})/P ~, \nonu \\
\alpha_{21} &=& (T_{21}T_{43}-T_{41}T_{23})/P ~, \nonu \\
\alpha_{22} &=& \lbrack(1-T_{22})P-(T_{21}+T_{23})(T_{32}+T_{12})
\rbrack /P ~,\nonu \\ 
P &=& T_{12}+T_{14}+T_{32}+T_{34} = T_{21}+T_{41}+T_{23}+T_{43} ~. \nonu \\
\eea
The general expression for the four-terminal resistance 
$R_{mn,kl}=1/G_{mn,kl}$ (which has six permutations) can then be written as 
\beq
R_{mn,kl} ~=~ \frac{h}{e^2} \frac{T_{km}T_{ln}-T_{kn}T_{lm}}{D} ~,
\eeq
where $D=(\alpha_{11}\alpha_{22}-\alpha_{12}\alpha_{21})/P$. From here, 
we can easily work out the four-terminal resistances 
for the case of the completely symmetric and maximally transmitting $S$-matrix 
(using Eq. (\ref{randt})). In this case however, as $\alpha_{12}
=\alpha_{21}=0$, all four-terminal resistances will simply give 
$R_{mn,kl}=0$. This result is interesting for the following reason: while
all the four-terminal resistances $R_{mn,kl}$ vanish if the system is 
exactly at the fully symmetric fixed point, the above suggests that they 
continue to stay zero as long as interactions are weak and the RG flows 
of the various elements of the $S$-matrix take place in a symmetric fashion. 
This means that in this case, the various $R_{mn,kl}$ will continue to be 
zero even as the temperature is varied.

The other interesting (and experimentally relevant) case for which 
conductances can be computed is that of two crossed, perfectly transmitting 
quantum wires which are connected via the tunneling of electrons at one point.
This point is characterized by the $S$-matrix given earlier in Eq. 
(\ref{smat3}). We have already seen that the hopping between the two wires is 
an irrelevant process. Further, we can treat any small reflection in either 
of the two perfectly transmitting wires perturbatively; from the work of 
Kane and Fisher \cite{kane}, it is known that such perturbations are relevant 
and will grow so as to cut the wires (i.e., they flow under RG towards the 
perfectly reflecting stable fixed point characterized by an $S$-matrix equal
to unity). Thus, there is nothing new to be found in the computation of
the conductances in this case.

Finally, we would like to mention the work of Komnik and Egger on crossed
quantum wires \cite{komnik}. In addition to the hopping operators 
considered above, they study the effects of a density-density interaction
between the two wires at the point where they cross; they show that such
an interaction can have a non-trivial effect if the interactions in the
wires are sufficiently strong. However, such strong interactions are beyond
the purview of our analysis; for the case of weak interactions considered 
here, such interactions are irrelevant.

\section{Tomonaga-Luttinger Liquids With Spin}

It is not difficult to extend all the results above to the case of interacting
fermions with spin. Let us first discuss the form of the interactions. We
again begin with a short-range interaction as in Eq. (\ref{hint1}) where
the density is now a sum of the form 
\beq
\rho ~=~ \Psi_\ua^\dagger \Psi_\ua ~+~ \Psi_\da^\dagger \Psi_\da ~. 
\eeq
The second-quantized
fields $\Psi_\ua$ and $\Psi_\da$ have expansions near the Fermi points
of the form given in Eq. (\ref{psiio}). (We assume that there is no magnetic
field, so that spin-$\ua$ and spin-$\da$ electrons have the same Fermi energy).
Following the arguments leading up to Eq. (\ref{hint2}), we can show that 
\bea
& & H_{int} = \nonu \\
& & \int dx \sum_{\sigma,\sigma^\prime =\ua,\da} [g_1
\Psi_{I\sigma}^\dagger \Psi_{O\sigma^\prime}^\dagger \Psi_{I\sigma^\prime} 
\Psi_{O\sigma}+g_2 \Psi_{I\sigma}^\dagger \Psi_{O\sigma^\prime}^\dagger 
\Psi_{O\sigma^\prime} \Psi_{I\sigma} \nonu \\ 
& & ~~~~~~~~~ + \frac{1}{2} g_4 ( \Psi_{I\sigma}^\dagger 
\Psi_{I\sigma^\prime}^\dagger \Psi_{I\sigma^\prime} \Psi_{I\sigma} + 
\Psi_{O\sigma}^\dagger \Psi_{O\sigma^\prime}^\dagger \Psi_{O\sigma^\prime} 
\Psi_{O\sigma} )], \nonu \\
& &
\label{hint3}
\eea
where 
\bea
g_1 ~&=&~ {\tilde V} (2k_F) ~, \nonu \\
{\rm and} \quad g_2 ~&=&~ g_4 ~=~ {\tilde V} (0) ~.
\eea
Yue et al show that the backscattering interaction governed by $g_1$ leads to 
a logarithmic 
renormalization of the interaction parameters $g_1$ and $g_2$ \cite{yue}; 
we will ignore that effect here since it plays no role to first order in the 
$g_i$. We can also ignore the effects of the $g_4$ term; it renormalizes the 
velocity, but it does not contribute to the reflection from the Friedel 
oscillations which is what leads to the RG flow of the $S$-matrix.

If there is a non-zero reflection amplitude $r$ on wire $i$, then there will
again be Friedel oscillations given by Eqs. (\ref{fried}-\ref{expec}) for 
both spin-$\ua$ and spin-$\da$ electrons. Then the interactions will lead to 
scattering of incoming electrons to outgoing electrons (and vice versa); this 
is given by the following Hartree-Fock decomposition of (\ref{hint3}),
\bea
& & H_{\rm int} ~= \nonu \\
& & - ~\frac{i(g_2 - 2 g_1)}{4\pi} \int_0^\infty dx ~[~ r^\star ~(
\Psi_{I\ua}^\dagger \Psi_{O\ua} ~+~ \Psi_{I\da}^\dagger \Psi_{O\da}) \nonu \\
& & ~~~~~~~~~~~~~~~~~~~~~~~~~~~~~~~ -~ r ~(\Psi_{O\ua}^\dagger \Psi_{I\ua} ~+~
\Psi_{O\da}^\dagger \Psi_{I\da}) ~]~ . \nonu \\
& &
\label{hf2}
\eea
(This may be compared with Eq. (\ref{hf1}) for spinless fermions). We see from
Eq. (\ref{hf2}) that the spin-$\ua$ and spin-$\da$ electrons have decoupled
from each other in this approximation. Hence the RG analysis given above for 
spinless fermions will go through similarly here. The only difference 
is that the interaction parameter $\alpha$ is now given by
\beq
\alpha ~=~ \frac{{\tilde V} (0) - 2 {\tilde V} (2k_F)}{2\pi \hbar v_F} ~,
\eeq
instead of $\alpha = [{\tilde V} (0) - {\tilde V} (2k_F)]/(2\pi \hbar v_F)$ 
in the spinless case. We thus see that to first order in the interaction, the 
analysis remains essentially the same for spinless and spinful fermions.
Finally, the conductances have factors of $2e^2 /h$ for spinful fermions
in place of $e^2 /h$ for spinless fermions but have similar temperature 
power-laws dependent on the interaction parameter $\alpha$ defined above. In 
fact, it should be possible to detect such power-laws in existing 
3-arm and 4-arm quantum wire systems built by the voltage-gate patterning 
on the 2DEG in GaAs heterojunctions \cite{timp,shepard}. While the early 
experiments with such systems focussed on carrier transport in the presence 
of an external magnetic field and the effects of geometry \cite{timp}, 
measuring the two-terminal, three-terminal and four-terminal conductances 
for fixed values of the various gate voltages but at different temperatures 
should again reveal identical power-law variations as discussed earlier for
spinless fermions. In fact, similar 
studies using the technique developed by Shepard {\it etal.} \cite{shepard} 
for directly measuring the transmission matrix elements of such junctions 
should be able to show the temperature power-law variations of the various 
transmission probabilities.

\section{\bf Conclusion}

In this work, we have derived the RG equations for a general $S$-matrix at 
the junction of several quantum wires, and we have discussed the consequences 
of these equations for the conductances across the system. The RG flows are a 
result of interactions in the wires; there is no flow if the interaction 
parameters $\alpha_i$ are all zero. Our results differ considerably from those
of Ref. \cite{nayak} who find RG flows even in the absence of interactions in 
the wires. This difference seems to be due to their model of the junction;
they have a spin-$1/2$ degree of freedom sitting there which interacts with
the electrons on the wires. This gives rise to a nontrivial interacting model 
of the Kondo type even if there are no interactions in the wires. Their flow 
diagram is therefore quite different from ours. Further, they only consider 
the case where both the $S$-matrix and the interactions are symmetric under 
all possible permutations of the wires; however they are able to use
bosonization to study the case of an arbitrary interaction strength.

Our work can clearly be generalized to the case of more than four wires
meeting at a junction. The RG flow diagram will rapidly get
more complicated as the number of wires increases. Physically, we expect 
the cases of three and four wires to be the easiest to study; these two cases 
arise in the experiments discussed earlier \cite{timp,shepard} as well as in
the cases of Y-branched carbon nanotubes \cite{papa} and crossed carbon 
nanotubes \cite{kim}. 

Besides the restriction to weak interactions in the wires, our work has
the limitation that we have assumed linear relations between the incoming 
and outgoing fermion fields. In principle, other interesting things 
can happen at a junction. For instance, there may be Andreev reflection in 
which a fermion striking the junction from one wire is reflected back as a 
hole while two fermions are transmitted into some of the other wires 
\cite{nayak}. Even more complicated things may
occur for the case of spinful fermions. Some of these phenomena can be
expressed as boundary conditions at the junction in the bosonic language, but
not in the fermionic language. We expect that such bosonic boundary conditions
will require a method of analysis which is very different from the one which
we have used to study the fermion $S$-matrix in this paper.

Finally, it remains a challenging problem to see if some of the non-trivial 
fixed points that we have found (such as case VII for the three-wire case)
can be bosonized for arbitrary interaction strengths. Bosonizing such points 
would lead to a much more complete picture of the RG flows
besides increasing our understanding of conformal field theories with 
boundaries.

\vskip .5 true cm
\leftline{\bf Acknowledgments}
\vskip .5 true cm

SR would like to acknowledge discussions with Ian Affleck, Cliff Burgess, Manu
Paranjape and Ashvin Vishwanath. DS thanks the Council of Scientific and 
Industrial Research, India for financial support through grant No. 
03(0911)/00/EMR-II.
\vskip .5 true cm

\end{document}